\documentstyle[12pt]{article}
\def\beq{\begin{equation}}
\def\eeq{\end{equation}}
\def\bea{\begin{eqnarray}}
\def\eea{\end{eqnarray}}
\begin {document}
\begin{titlepage}
March 1996 \\
\begin{flushright}
HU Berlin-EP-96/8\\
\end{flushright}
\mbox{ }  \hfill hepth@xxx/9603186
\vspace{6ex}
\Large
\begin {center}
\bf{On T-duality for open strings in general abelian and nonabelian
gauge field backgrounds}
\end {center}
\large
\vspace{3ex}
\begin{center}
H. Dorn, H.-J. Otto
\footnote{e-mail: dorn,otto@qft3.physik.hu-berlin.de}
\end{center}
\normalsize
\it
\vspace{3ex}
\begin{center}
Humboldt--Universit\"at zu Berlin \\
Institut f\"ur Physik, Theorie der Elementarteilchen \\
Invalidenstra\ss e 110, D-10115 Berlin
\end{center}
\vspace{6 ex }
\rm
\begin{center}
\bf{Abstract}
\end{center}
We discuss T-duality for open strings in general background fields both in 
the functional integral formulation as well as in the language of canonical 
transformations. The Dirichlet boundary condition in the dual theory has to 
be treated as a constraint on the functional integration. Furthermore, we 
give meaning to the notion of matrix valued string end point position in the 
presence of nonabelian gauge field background.
\vspace{3ex}
\end{titlepage}
%%%%%%%%%%%%%%%%%%%%
\setcounter{page}{1}
\pagestyle{plain}
\section{Introduction}
Target space duality and its implications for string theory have been studied
to a great extent and in many details, see \cite{givrev, alvg} and
references therein. All these investigations refer to closed strings.
However, due to the crucial observation \cite{pol2} that Dirichlet branes
\cite{pol1} play a prominent role in the general duality pattern
of string theory there is now a considerable activity on these D-branes
including aspects of their T-duality relations to open strings.

Generalized $\sigma $-models on 2D manifolds with boundary describing
open strings in the background of target space fields corresponding to
excitations of both closed and open strings are the natural framework to
proceed from strings in flat (and in some directions compactified) target
space to the more general situation of nontrivial backgrounds and to derive
the open string extension of Buscher's formulae \cite{busch}. Besides the
early paper \cite{leigh} this question has been addressed only very recently
\cite{alv}. \footnote{A preliminary version of our paper has been reported
by H.D. at the workshop ``Physics beyond the standard model'', Bad Honnef,
March 11-14. While preparing this manuscript ref. \cite{alv} appeared with
partial overlap in the abelian case.}

Although the basic new effect of T-duality in the open string case, the
mapping of Neumann to mixed Neumann $and$ Dirichlet boundary conditions
follows immediately from the closed string duality transformation
$\partial _m X\rightarrow \epsilon _{mn}\partial _n X$, a more careful analysis
is necessary to uncover the precise meaning of the boundary condition
in the quantized 2D world sheet field theory. At the classical level it makes
no difference whether the boundary condition is treated as an external
constraint on the fields competing in the action principle or whether
these fields are varied freely on the boundary and the boundary condition
arises as part of the stationarity condition of the action on the same
footing as the equation of motion in the bulk. However, in the quantized
theory both procedures may yield different theories.

Our aim  is to  contribute to a systematic study of the 2D field theoretical
aspects of the boundary in T-duality. We will investigate a model containing
target space fields corresponding to the metric $G$, the antisymmetric tensor
$B$, the dilaton $\Phi $, a gauge field $A$ and a technically motivated
field $V$ \cite{leigh}. The model can be used to deal with the open oriented
bosonic string ($B\neq 0$, gauge group U(N)) or with the type I superstring
in nonvanishing bosonic backgrounds ($B=0$, gauge group SO(N)). Only
local world sheet effects will be considered.

In section 2 we treat the case of an abelian gauge field via functional
integral manipulations. Section 3 presents a formalism to handle
nonabelian gauge fields thereby giving meaning to the notation
of open strings bound with their end points to a matrix D-brane. Section 4
supplements the functional integral treatment by a look at the implications
of the presence of a boundary on the interpretation of T-duality as a
canonical transformation.
%%%%%%%%%%%%%%
\section{Functional integral treatment of T-duality}
Our starting point is a generalized $\sigma$-model action describing a
string moving in background fields corresponding to massless excitations
of both closed ($\psi =(G_{\mu \nu},~ B_{\mu \nu},~\Phi)$)and open string
states (gauge field $A_{\mu}$, for simplicity first of abelian type)
\bea
&S&[\psi ,A,V;X]~=~\frac{1}{4\pi \alpha ^{\prime}}\int _M d^2 z \sqrt{-g}
\left ( \partial _m X^{\mu} \partial _n X^{\nu} E^{mn}_{\mu \nu}+
\alpha ^{\prime} R^{(2)}\Phi (X) \right )
\nonumber \\
&+&\int _{\partial M} \left (A_{\mu}(X(z(s)))\partial _nX^{\mu}(z(s))\dot{z}^n
+V_{\mu}\partial _m X^{\mu}\frac{\epsilon ^{mn}}{\sqrt{-g}}\dot{z}_n (s)-\frac
{1}{2\pi}k(s)\Phi \right )ds~,\nonumber \\
&&E^{mn}_{\mu \nu}~=~g^{mn}G_{\mu \nu}(X)~+~\frac{\epsilon ^{mn}}{\sqrt{-g}}
B_{\mu \nu}(X)~.
\label{a}
\eea
$R^{(2)}$ is the 2D curvature scalar, $k(s)$ the geodesic curvature on the
boundary $\partial M$. We added a boundary coupling of the normal
derivative of the world sheet position to a vector field $V$. This field does
not correspond to any string excitation. It is added for technical reasons.
Even starting with $V=0$ there appear counter terms proportional
to the normal derivative \cite{do}. Although for free string ends they can be
included in the $A$ renormalization by using the boundary condition \cite{cal1}
they get independent physical meaning in the presence of Dirichlet boundary
conditions \cite{leigh}. In the rest of the paper we take flat 2D metrics and
disregard the dilaton field $\Phi$ whose transformation under T-duality should
arise from some functional Jacobian like in the closed string case
\cite{busch}.

We assume the existence of one Killing vector field $k^{\mu}(X)$ and invariance
of our action under a diffeomorphism in the direction of $k$
\footnote{Having in mind the nonabelian generalization we ignore the 
possibility ${\cal L} _kB=-{\cal L}_kF\neq 0$.}
\beq
{\cal L}_k\psi=0,~ {\cal L}_kV=0,~ {\cal L}_kF=0~.
\label{b}
\eeq
The condition on the field strength $F$ implies
\beq
{\cal L}_kA_{\mu}=\partial _{\mu}a~.
\label{c}
\eeq
Choosing suitable adapted coordinates we have
\beq
X^{\mu}=(X^0,X^M),~~~k^{\mu}=(1,0),~~~
{\cal L}_k=\partial _0~.
\label{d}
\eeq
Starting from a generic allowed gauge field fulfilling $\partial _0 A_{\mu}=
\partial _\mu a$ one can reach 
\beq
\partial _0 A_{\mu}~=~0
\label{e}
\eeq
via a gauge transformation. We take (\ref{e}) as a gauge condition. 
Altogether now all our target space fields in (\ref{a}) are independent of 
$X^0$. For later use we classify the residual gauge transformations 
respecting (\ref{e}). The obvious condition $\partial _0\partial_{\mu}\Lambda
=0$ implies $\Lambda (X^{\mu})=\lambda (X^M)+c\cdot X^0$, i.e. the still 
allowed gauge transformations are
\beq
A_0\rightarrow A_0+c~,~~~~~A_M\rightarrow A_M+\partial _M \lambda (X^M)~.
\label{f}
\eeq

To start the dualization procedure we rewrite the partition function
\footnote{ To identify the dual target space fields one can restrict
oneselves to the partition function. A more complete study of the
2D field theory would require the inclusion of source terms for $X^{\mu}$.}
of our $\sigma$ model in standard way with the help of a Lagrange multiplyer
field $\tilde{X}^0$:
\beq
Z[\psi ,A,V]~=~\int DX^{\mu}e^{iS[\psi ,A,V;X^{\mu}]}~=~
\int DX^{M}Dy_mD\tilde{X}^0 e^{i\bar{S}[\psi , A,V;X^M,y_m,\tilde{X}^0]}~,
\label{g}
\eeq
with
\bea
\bar S [\psi ,A,V;X^M,y_m,\tilde{X}^0]&=&\frac{1}{4\pi \alpha ^{\prime}}
\int _M d^2 z( \partial _m X^{M} \partial _n X^{N} E^{mn}_{MN}+
2\partial _mX^M y_n E^{mn}_{M0}\nonumber \\
&&~~~~~~~~~~~~~~~~~+y_m y_n E^{mn}_{00}+2\tilde{X}^0\epsilon ^{mn}
\partial _m y_n)\label{h}\\
&+&\int _{\partial M} (A_M \partial _n X^M+A_0y_n+(V_M
\partial _m X^M+V_0y_m)\epsilon ^{m~}_{~~n})\dot{z}^n (s)ds~.
\nonumber
\eea
The dualized version will be obtained by performing the $y_m$ functional
integration. In the closed string case this integral decouples after a shift
read off from a suitable quadratic completion. Now the presence of the
boundary induces a crucial subtlety:
\bea
\bar S&=&\frac{1}{4\pi \alpha ^{\prime}}\int _M d^2 z\left (g^{mn}G_{00}
v_mv_n+E^{mn}_{MN}\partial _m X^{M} \partial _n X^{N}-\frac{g^{mn}}{G_{00}}
K_m K_n \right )\nonumber \\
&+&\int _{\partial M}(A_M \partial _n X^M+V_M\partial _m X^M\epsilon ^{m~}_
{~~n})\dot{z}^n ds+ \int _{\partial M}\left (v_m L^m-\frac{1}{G_{00}}K_m L^m
\right )ds~,
\label{i}
\eea
with
\bea
K^m&=&E^{mn}_{0N}\partial _n X^M+\epsilon ^{mn}\partial _n\tilde{X}^0~,
\nonumber\\
L^m&=&A_0\dot{z}^m+V_0\epsilon ^{m~}_{~~n}\dot{z}^n+\frac{1}{2\pi
\alpha ^{\prime}}\tilde X^0\dot{z}^m~,~~~~~~~~~~~~v_m=y_m+\frac{K_m}{G_{00}}~.
\label{j}
\eea
Due to the dependence of $L^m$ on $X^M,~\tilde X ^0$ the $v$ integral even
after the rescaling $v\rightarrow \sqrt{G_{00}}v$ does not decouple.

Since the $v$ integral is ultralocal we can use
\beq
\int _M Dv_me^{i\bar S}~=~\int _M Dv_m\exp \left (i\bar S-i\int _{\partial M}
v_mL^mds\right )\cdot \int _{\partial M}Dv_m\exp \left (i\int _{\partial M}
v_mL^mds\right )~.
\label{k}
\eeq
Then the $v_m$ integral on the boundary $\partial M$ produces a functional
$\delta$ function which imposes for the remaining functional integral the
constraints $L^m(s)=0,~~m=1,2$, i.e.
\beq
V_0(X^M)~=~0~,
\label{l}
\eeq
\beq
2\pi \alpha ^{\prime}A_0 (X^M)~+~\tilde X^0~=~0~~~~\mbox{on}~\partial M~.
\label{m}
\eeq
Including the remaining bulk $v_m$ integral into the overall normalization
we finally arrive at
\beq
Z[\psi ,A,V]~=~\int _{(\ref{l}),(\ref{m})}D \tilde X^{\mu}\exp (iS[\tilde
{\psi},\tilde A,\tilde V;\tilde X])~,
\label{n}
\eeq
with
\beq
\tilde X^{\mu}~=~(\tilde X^0,X^M)~,
\label{o}
\eeq
\beq
\tilde A_{\mu}~=~(0,A_M)~,~~~~~\tilde V_{\mu}~=~(0,V_M)~,
\label{p}
\eeq
\bea
\tilde G_{00}=\frac{1}{G_{00}}~,~~~~~\tilde G_{0M}=\frac{B_{0M}}{G_{00}}~,
~~~~~\tilde G_{MN}=G_{MN}-\frac{G_{M0}G_{0N}+B_{M0}B_{0N}}{G_{00}}~,
\nonumber\\
\tilde B_{0M}=\frac{G_{0M}}{G_{00}}~, ~~~~~\tilde B_{MN}=B_{MN}-\frac
{G_{M0}B_{0N}+B_{M0}G_{0N}}{G_{00}}~.
\label{q}
\eea
The formulas (\ref{q}) coincide with that from the closed string case
\cite{busch}. The notation in (\ref{n}) means that the integrand of
the functional integral has to respect the boundary conditions (\ref{l})
and (\ref{m}). In general the $two$ boundary conditions are
not compatible and we have to put $V\equiv 0$ \cite{leigh}.
Our model in its dual version describes an open string bound with its end
points to the D-brane $\tilde X^0=-2\pi\alpha ^{\prime}A_0(\tilde X^M)$.
The boundary condition arising in the course of T-dualization appears as an 
external condition on the quantum theory, it is not only a consequence of the 
stationarity condition of the action.
Due to (\ref{f}) gauge invariance in the original theory is mapped into
gauge invariance for the gauge fields parallel to the D-brane and to
translational invariance of the D-brane position in the dual theory.
%%%%%%%%%%%%%%%%%%%%%%%
\section{Open strings with nonabelian Chan-Paton factors}
We begin with the partition function for the generalized $\sigma$-model
describing an open string~\footnote{For notational simplicity we drop $V$.}
in nontrivial metric, antisymmetric tensor,
dilaton and nonabelian gauge field background \cite{ts, cal2, do}
\beq
Z[\psi ,A]~=~\int DX~e^{iS[\psi,A=0,V=0;X]}\mbox{tr}Pe^{i\int _
{\partial M}A_{\mu}dX^{\mu}}~.
\label{1}
\eeq
This expression no longer has the standard form of a functional integral
over the exponential of a local action. We can achieve such a structure by
the introduction of a one-dimensional auxiliary field $\zeta _a (s)$ living
on the boundary $\partial M$. It has the propagator $\langle \bar{\zeta}_a
(s_1)\zeta _b (s_2)\rangle _0=\delta _{ab}\Theta (s_2-s_1)$ and a coupling
to $X^{\mu}$ via the interaction term \cite{zform} 
$$i\bar{\zeta}A_{\mu}(X(z(s)))\zeta (s)\partial _m X^{\mu}\dot{z}^m(s)$$ 
($z(s)$ parametrizes $\partial M,~0\leq s\leq 1$)~\footnote{For a similar 
but slightly different formalism
see ref. \cite{ms}.}.
\beq
Z=\int DX~D\bar{\zeta}~D\zeta ~\bar{\zeta}_a(0)\zeta _a(1)~
e^{iS_0[\bar{\zeta}, \zeta ]}\exp (i\hat{S}[\psi ,\bar{\zeta}A_{\mu}
\zeta ;X]).
\label{2}
\eeq
For general gauge groups there is an additional factor induced by the
vacuum graphs of the $\zeta$-theory. The form presented in (\ref{2}) is
valid for unimodular Wilson loops only, e.g. elements of SU(n) or SO(n).
In this case the choice $\zeta$ fermionic or bosonic is irrelevant.
To condense the notation we introduced an action $\hat S$ depending on the
target space fields $\psi = (G,~B,~\Phi )$ the string world sheet position
$X^{\mu}(z)$ and a vector field $C_{\mu}(X^M,s)$ with possibly explicit 
dependence on $s$ by
\bea
\hat{S}[\psi ,C;X]&=&\frac{1}{4\pi \alpha ^{\prime}}\int _M d^2 z \sqrt{-g}
\left ( \partial _m X^{\mu} \partial _n X^{\nu} E^{mn}_{\mu \nu}(X(z))+
\alpha ^{\prime} R^{(2)}(z) \Phi (X(z)) \right )
\nonumber \\
&+&\int _{\partial M} \left (C_{\mu}(X(z(s)),s)\dot X^{\mu}-\frac{1}{2\pi}
k(s)\Phi (X(z(s))) \right )ds~.
\label{3}
\eea
Under the $\zeta$-path integral we now can repeat the dualization procedure
\footnote{Assuming again independence of $X^0$ for all target space fields
involved.}
performed above for the case of abelian $A$. Due to the presence of
$\zeta (s)$ in $C_{\mu}$ the resulting Dirichlet condition, besides its
implicit dependence on $s$ via the embedding of the world sheet boundary in
the target space, also depends on $s$ explicitly. Introducing
\beq
{\cal F}[\psi ,C\vert f]~=~\int _{X^0(z(s))=f(X^M(z(s)),s)} D X^{\mu}~\exp 
(i\hat{S}[\psi ,C;X])
\label{4}
\eeq
we can write our partition function $Z$ as
\beq
Z[\psi,A]~=~ \int D\bar{\zeta}~ D\zeta ~\bar{\zeta}_a (0)\zeta _a (1)e^{iS_0 [
\bar{\zeta},\zeta ]}~{\cal F}[\tilde{\psi},\bar{\zeta}
\tilde A \zeta \vert -2\pi\alpha ^{\prime}\bar{\zeta}A_0\zeta ]~.
\label{5}
\eeq
The $\zeta $-integration results in ordering the matrices sandwiched between
$\bar{\zeta}(s)$ and $\zeta (s)$ with respect to increasing $s$.
But unfortunately after performing the functional integral over the world 
surfaces $X^{\mu}(z)$ there is no longer any correlation between $s$ and the
target space coordinate at which the gauge field has to be taken. The
arguments of ${\cal F}$ are $\tilde{\psi}(x^M)$, $C(x^M,s)=\bar{\zeta}(s)
\tilde A (x^M)\zeta (s)$ and $f(x^M,s)=\bar{\zeta}(s)A_0(x^M)\zeta (s)$.
\footnote{We use $x^{\mu}$ to denote the target space coordinates in
writing $\tilde{\psi} $ and $\tilde A$ as target space functions. A way to 
disentangle $x^{\mu}$ and the functional integration variable $X^{\mu}$
is to write e.g. $\tilde{\psi}(X)=\int dx~\tilde{\psi}(x)\delta (x-X(z))$ 
\cite{tsdelt}.}  
Therefore, we cannot further simplify (\ref{5}) by manipulations with
the general formal expression.

However, usually all calculations in the $\sigma $-model framework are done 
within the background field method. We assume that its justification along the
line of arguments of \cite{hopast} can be extended to the case of manifolds
with boundary and the type of boundary conditions involved in our above 
discussion. Then the background field method variant of (\ref{4},\ref{5}) 
with ($\xi $ denoting Riemann normal coordinates in the target space)
\beq
X^{\mu}(z)~=~X^{\mu}_{cl}(z)~+~\delta X^{\mu}(\xi (z))
\label{A}
\eeq
is
\bea
&(\ast )&~~~~~~X^0(z(s))=f(X^M(z(s)),s)\nonumber\\[3mm]
\lefteqn{{\cal F}_{bm}[\psi (X_{cl}),C(X_{cl},s);X_{cl}\vert f]~=}&&
\label{B}\\
&&\int _{(\ast )}D \xi ^{\mu}~\exp(i\hat{S}[\psi (X_{cl}+\delta X(\xi)),C(X_{cl}+\delta X
(\xi),s);X_{cl}+\delta X(\xi )])
\nonumber
\eea
and
\bea
\lefteqn{Z_{bm}[\psi (X_{cl}),A(X_{cl});X_{cl}]~=}\label{C}\\[2mm]
&& \int D\bar{\zeta}~ D\zeta ~\bar{\zeta}_a (0)\zeta _a (1)e^{iS_0 [
\bar{\zeta},\zeta ]}~{\cal F}_{bm}[\tilde{\psi},\bar{\zeta}
\tilde A \zeta ;\tilde X_{cl}\vert -2\pi\alpha ^{\prime}\bar{\zeta}A_0\zeta ]~.
\nonumber
\eea
%%%%%%%%%%%%%%%%%%%%%%%%

The functional ${\cal F}_{bm}$ depending on the Dirichlet boundary 
function $f(s)$ 
at first sight introduces a considerable generalization if explicit
$s$-dependence is allowed. Then the target space interpretation of a D-brane
with attached open string breaks down. However, we have to remember
that afterwards we need the special case $f(s)=-2\pi \alpha ^{\prime}
\bar{\zeta}A_0\zeta$ only.
In the perturbative evaluation of the $\zeta$ path integral the $\zeta$'s
combine to propagators which are nearly constant, either $0$ or $1$ in a way
just realizing path ordering of the accompanying $A$-matrices. Although
the original definition of ${\cal F}_{bm}$ in (\ref{B}) makes sense for scalar (not
matrix valued) $f(s)$ only, at least if ${\cal F}_{bm}$ has a well defined 
functional Taylor expansion in $f$ one can put into $this$ expansion also a 
matrix-valued $f$. The statement that the $\zeta$-integral results in nothing 
else but the path ordering can be proven by the following step (until the final
formula we suppress a factor $2\pi\alpha ^{\prime}$)
\bea
Z_{bm}&=&\left .{\cal F}_{bm}[\tilde{\psi},\frac{\delta}{\delta J};\tilde X
_{cl}\vert \frac{\delta}{\delta K} ]\int D\bar{\zeta}~ D\zeta ~\bar{\zeta}_a
(0)\zeta _a (1)e^{iS_0 +\int (\bar{\zeta}
\tilde A_M \zeta J^M-\bar{\zeta}A_0 \zeta K))ds}
\right | _{J=0,K=0}\nonumber \\
&=&\left . {\cal F}_{bm}[\tilde{\psi},\frac{\delta}{\delta J}
;\tilde{X}_{cl}\vert \frac{\delta}{\delta K} ]\mbox{tr}Pe^{\int (\tilde A_M 
J^M-A_0 K)ds}\right | _{J=0,K=0}~.
\label{6}
\eea
Since under the path ordering differentiation acts as if all quantities
would commute we arrive at (note (\ref{o},\ref{p}))
\beq
Z_{bm}[\psi ,A;X_{cl}]~=~ \mbox{tr}P({\cal F}_{bm}[\tilde{\psi},\tilde A;
\tilde X_{cl}\vert -2\pi 
\alpha ^{\prime}A_0])~.
\label{7}
\eeq
Together with eq.(\ref{B}) this yields an explicit realization of the statement
that in the nonabelian case the end point position of the open string in the
dual prescription becomes a matrix \cite{wit, polrev}.

Before any further interpretation we should have a closer look on the
consequences of gauge invariance. As in the abelian case in a generic
situation the Lie derivative of the gauge field can be nonzero.
${\cal L}_kA_{\mu}=D_{\mu}a$ ensures ${\cal L}_k F_{\mu \nu}=-i[F_{\mu\nu},
a]$ and henceforth the invariance of gauge invariant quantities like
tr$(F_{\mu\nu}F^{\mu\nu})$. However, by a suitable gauge transformation
again one can reach ${\cal L}_k A_{\mu}=0$. Hence we are allowed to enforce
in adapted coordinates (\ref{d}) the gauge condition (\ref{e}). The condition
constraining the allowed residual (infinitesimal) gauge transformations
is $D_{\mu}\partial _0 \Lambda = 0$. If a Lie algebra valued field $B(X)$
is covariantly constant in a whole neighborhood of a point $X$ its value at
any point $Y$ in this neighborhood  must fulfill the equation
\beq
B(Y)=P\exp \left (i\int _{\cal C}A_{\mu}dX^{\mu}\right )B(X)P\exp \left
(i\int _{{\cal C}^{-1}}A_{\mu}dX^{\mu}\right )
\label{8}
\eeq
for arbitrary curves ${\cal C}$ connecting $X$ and $Y$. Comparing (\ref{8})
for two different curves we conclude that $B(X)$ has to commute with
all path ordered phase factors for closed curves starting and ending at $X$,
i.e. with all elements of the holonomy group. The most restrictive situation
arises if this group coincides with or is a irreducible subgroup of the gauge
group. Then $B$ must be a multiple of unity. As a consequence 
$B(X)$ commutes with $A_{\mu}(X)$ and therefore covariant constancy implies
constancy. Applying this to $\partial _0\Lambda $ we get
$\Lambda = cX^0\cdot {\bf 1}+\lambda (X^M)$ and for the residual gauge
transformations the nonabelian version of (\ref{f})
\beq
A_0\rightarrow A_0+c\cdot {\bf 1}-i[A_0 ,\lambda (X^M)]~,~~~~~A_M\rightarrow
A_M+D_M \lambda (X^M)~.
\label{9}
\eeq
Of course for less restrictive holonomy groups there are more general
gauge transformations left.
Translational invariance of the D-brane position as a consequence of
gauge invariance in the original model one has only for gauge groups
where $c{\bf 1}$ is an element of the corresponding Lie algebra, e.g.
for U(N). For SO(N) $c$ has to vanish. This breaking of translational
invariance is in accord with the orientifold construction \cite{polrev}.

Furthermore, the transformation (\ref{9}) can always be used to bring
the matrix $A_0$ into the simplest possible form: diagonal for U(N) and
block diagonal (with 2x2 blocks) for SO(N).

Discussing string endpoints with definite Chan-Paton indices corresponds to
the situation $before$ the $\zeta$-integral is performed. Our partition 
function can be understood as an off shell extension of the generating 
functional of S-matrix elements for the elementary string excitations. 
Therefore,
the boundary of our world sheet $\partial M$ carries two Chan-Paton indices
related to the two indices associated with the endpoints of the would be
emitted open string state. Choosing $\bar{\zeta}_a=\delta _{ab},~\zeta_a=
\delta _{ac}$ one has $\bar{\zeta}A_0\zeta=(A_0)_{bc}$ and the boundary point
under consideration has to be on the hypersurface defined by $\tilde{X}_0
=-2\pi\alpha ^{\prime}(A_0)_{bc}$. Specializing to the case of diagonal $A_0$
we reproduce the brane pattern obtained in the presence
of Wilson lines \cite{polrev}.
%%%%%%%%%%%%%%%
\section{T-duality as a canonical transformation}
For a general 2D field theory on a manifold with boundary with the action
\beq
S[\phi ]~=~\int _M d^2zL(\phi , \partial\phi )~+~\int _{\partial M}
ds~l(\phi , \partial\phi )
\label{10}
\eeq
stationarity under free variation of $\phi $ both inside $M$ and on
$\partial M$ implies besides the standard bulk equation of motion
\bea
\frac{\partial L}{\partial\partial _a\phi}\epsilon _{ab}\dot z^b~+~\frac
{\partial l}{\partial\phi}~-~\partial _{{\bf t}}\frac{\partial l}
{\partial _{{\bf t}\phi}}~=~0~,\nonumber \\
\frac{\partial l}{\partial \partial _{{\bf n}}\phi}~=~0~.
\label{11}
\eea
The indices ${\bf t}$ and ${\bf n}$ denote tangential and normal directions,
respectively. In general it may happen that the two boundary conditions arising
in (\ref{11}) are not compatible.

For the transition to the Hamilton formulation we denote the 1D space
coordinate by $\sigma $ and the time coordinate by $\tau $. Let us assume
that the boundary consists of two parts $\partial M\vert _i$ and $\partial
M\vert _f$ which are identified in the infinite past and future. To avoid
distributional formulas we do not include the boundary action in the
definition of canonical momenta $\pi (\sigma ) =\partial L/\partial
\dot{\phi}(\sigma )$. Then the Hamilton function is of the Routh type in so 
far as the boundary degrees of freedom are kept in Lagrangian form. We skip
writing down the resulting equations and proceed directly to the canonical
transformations keeping them form invariant. For a generating function of the
type $F(\phi ,\tilde{\phi})$ we get
\bea
\tilde{\pi}&=&-\frac{\partial F}{\partial\tilde{\phi}}+\partial _
{\sigma}\frac{\partial  F}{\partial \tilde{\phi}'}~,~~~~~
\pi~=~\frac{\partial F}{\partial \phi}-\partial _{\sigma}
\frac{\partial F}{\partial \phi '}~,~~~~~
\tilde H~=~H+\frac{\partial F}{\partial \tau}~,\nonumber\\
\tilde l&=&l\pm \frac{\partial F}{\partial \phi '}\dot{\phi}\pm
\frac{\partial F}{\partial\tilde{\phi}'}\dot{\tilde{\phi}}~.
\label{12}
\eea
The $\pm$ alternative refers to the two boundary components corresponding
to the two string end points. The dot and prime denote differentiation with
respect to $\tau$ and $\sigma$, respectively.

We now apply this general framework to our $\sigma $-model with one
abelian target space isometry. The generating function
\beq
F~=~\frac{1}{4\pi \alpha ^{\prime}}\left (X^{\prime\, 0}\tilde{X}^0~-~X^0\tilde
{X}^{\prime\, 0}
\right )
\label{13}
\eeq
known already from the closed string case \cite{alvg} leads with (\ref{12})
to
\beq
\dot X^0~=~-\frac{\tilde X^{\prime\, 0}+G_{0N}\dot X^N+B_{0N}X^{\prime\, N}}{G_
{00}}
\label{14}
\eeq
and
\beq
\pm \tilde l~=~A_M\dot X^M~-~ \left (A_0+\frac{\tilde X^0}{2\pi\alpha ^
{\prime}}
\right )\cdot\frac{\tilde X^{\prime\, 0}+G_{0N}\dot X^N+B_{0N}X^{\prime\, N}}
{G_{00}}~.
\label{15}
\eeq
Although we started in the original theory with $V\equiv 0$ the dualized
version contains a boundary interaction of the $V$-type, i.e.
proportional to the normal derivative of $\tilde X^0$.~\footnote{Note that we 
have choosen the $\sigma$-$\tau$ coordinates in such a
way that on the boundary $\partial /\partial \sigma $ is the normal
and $\partial /\partial\tau $ the tangential derivative.} The boundary
condition arising from $\partial\tilde l /\partial \tilde X^{\prime\, 0} =0$ is
given by (\ref{m}). Using this the other conditions from (\ref{11})
applied to the dual theory yield Neumann conditions for the remaining
target space coordinates $\tilde X^M=X^M$. Since Neumann and Dirichlet
refer to different coordinates there is no compatibility problem.
%%%%%%%%%%%%%%%
\section{Conclusions}
We have sketched how the well known techniques to describe T-duality
for closed strings in general background fields can be extended to the case of
open strings. In the simplest case of one abelian isometry the dual theory
corresponds to strings bound with their ends to the hypersurface whose
position in target space is given by setting the coordinate in the direction 
of the Killing vector equal to the corresponding component of the gauge 
potential. In the language of canonical transformations we see no simple
possibility to decide whether the boundary condition is a dynamical one,
resulting as part of the stationarity condition for the action, or whether
it is an external constraint. On the other hand the functional integral 
treatment clearly shows that the Dirichlet condition has to be handled as a 
constraint on the integrand of the functional integral.

Using the $\zeta$ auxiliary field formalism we were able to proceed from
abelian gauge fields to nonabelian ones. Here at a first step one has to
calculate the partition function for explicitly boundary parameter dependent
Dirichlet conditions and then, in a second step, to insert in the arising 
functional the matrix valued gauge potential in a path ordered manner.

Obviously, a lot of interesting problems remain open for further 
investigations. One should clarify the structure of residual gauge
transformations in the case of holonomy groups not as restrictive as assumed
in (\ref{9}). Furthermore, the relation between $\beta $-functions in the
original and dual theory and their role in deriving equations of motion
for the D-brane discussed in ref. \cite{leigh} for the abelian case has to
be extended to nonabelian gauge fields. Concerning more general aspects the
inclusion of fermions, questions of supersymmetry and global issues
\cite{alv} require further study.\\[5mm] 
%%%%%%%%%%%%%%%%
{\bf Acknowledgement}\\ 
We thank K. Behrndt, S. F\"orste, A. Kehagias, T. Mohaupt and C. Preitschopf 
for useful discussions.
%%%%%%%%%%%%%%%%%


\begin{thebibliography}{77}
\bibitem{givrev}
A. Giveon, M. Poratti, E. Rabinovici, {\it Phys. Rep.} {\bf 244} (1994) 77

\bibitem{alvg}
E. Alvarez, L. Alvarez-Gaum\'e, Y. Lozano, {\it Nucl. Phys.} {\bf B41},
Proc. Suppl. (1995) 1

\bibitem{pol2}
J. Polchinski, {\it Phys. Rev. Lett.} {\bf 75} (1995) 4724

\bibitem{pol1}
J. Dai, R.G. Leigh, J. Polchinski, {\it Mod. Phys. Lett.} {\bf A4} (1989)
2073\\
P. Horava, {\it Phys. Lett.} {\bf B231} (1989) 251\\
M.B. Green, {\it Phys. Lett.} {\bf B266} (1991) 325

\bibitem{busch}
T.H. Buscher, {\it Phys. Lett.} {\bf B194} (1987) 51,
              {\it Phys. Lett.} {\bf B201} (1988) 466

\bibitem{leigh}
R.G. Leigh, {\it Mod. Phys. Lett.} {\bf A4} (1989) 2767

\bibitem{alv}
E. Alvarez, J.L.F. Barbon, J. Borlaf, {\it T-duality for open strings}
preprint hep-th/9603089

\bibitem{do}
H. Dorn, H.-J. Otto, {\it Zeitschr. f. Phys.} {\bf C32} (1986) 599

\bibitem{cal1}
C.G. Callan, C. Lovelace, C.R. Nappi, S.A. Yost, {\it Nucl. Phys.} {\bf B288}
(1987) 525

\bibitem{ts}
E.S. Fradkin, A.A. Tseytlin, {\it Nucl. Phys.} {\bf B261} (1985) 1\\
A.A. Tseytlin, {\it Nucl. Phys.} {\bf B276} (1986) 391

\bibitem{cal2}
C.G. Callan, E. Martinec, M.J. Perry, D. Friedan, {\it Nucl.Phys.}
{\bf B262} (1985) 593

\bibitem{zform}
S. Samuel, {\it Nucl. Phys.} {\bf B149} (1979) 517\\
R.A. Brandt, F. Neri, D. Zwanziger, {\it Phys. Rev.} {\bf D19} (1979) 1153\\
J.L. Gervais, A. Neveu, {\it Phys. Rev.} {\bf B163} (1980) 189\\
I. Ya. Arefyeva, {\it Phys. Lett.} {\bf 93B} (1980) 347\\
H. Dorn, {\it Fortschr. d. Phys.} {\bf 34} (1986) 11

\bibitem{ms}
N. Marcus, A. Sagnotti, {\it Phys. Lett.} {\bf 188B} (1987) 58

\bibitem{wit}
E. Witten, {\it Nucl. Phys.} {\bf B460} (1996) 335

\bibitem{polrev}
J. Polchinski, S. Chaudhuri, C.V. Johnson, preprint {\it Notes on D-Branes},
hep-th/9602052

\bibitem{tsdelt}
A.A. Tseytlin, {\it Nucl. Phys.} {\bf B294} (1987) 383

\bibitem{hopast}
P.S. Howe, G. Papadopoulos, K.S. Stelle, {\it Nucl. Phys.} {\bf B296} (1987)
26

\end{thebibliography}
\end{document}